\documentclass[acmsmall]{acmart}
\AtBeginDocument{%
  }

\acmConference[FSE '26]{
the ACM International Conference on the Foundations of Software Engineering}{July 5--9, 2026}{Montreal, Canada}

\pdfmapfile{=pdftex.map}
\usepackage{algorithmic}
\usepackage{graphicx}
\usepackage{textcomp}
\usepackage{xcolor}
\usepackage{pgfplots}
\usepackage{listings}
\usepackage{color,soul}

\newcommand{\algorithmautorefname}{Algorithm}
\newcommand{\refappendix}[1]{\hyperref[#1]{Appendix~\ref*{#1}}}
\def\sectionautorefname~#1\null{Section #1\null}
\def\equationautorefname~#1\null{Equation (#1)\null}
\def\algorithmautorefname~#1\null{Algorithm (#1)\null}
\def\subsectionautorefname~#1\null{Section #1\null}
\def\subsubsectionautorefname~#1\null{Section #1\null}

\usepackage{multirow}
\usepackage{kotex}
\usepackage{url}
\usepackage{xspace}

\usepackage{xstring}
\newcommand{\PP}[1]{
	\vspace{4px}
\noindent{\bf \IfEndWith{#1}{.}{#1}{#1.}}}

\newcommand{\etal}{\textit{et al}.\xspace}
\newcommand{\ie}{\textit{i}.\textit{e}.}
\newcommand{\eg}{\textit{e}.\textit{g}.}
\usepackage{subcaption}

\usepackage[skins]{tcolorbox}
\tcbuselibrary{breakable}
\lstdefinestyle{base}{
	language=C,
	numbers=left,
	numberstyle=\tiny,
	numbersep=-3pt,
	breaklines=true,
	commentstyle=\color{mygray},
	columns=fullflexible,
	basicstyle=\fontsize{7.3}{9}\ttfamily\color{black},
	moredelim=**[is][\bfseries]{@bold-}{-bold@},
	moredelim=**[is][\color{tomato}]{@R-}{-R@},
	moredelim=**[is][\color{mygreen}]{@G-}{-G@},
	moredelim=**[is][\color{mygray}\bfseries]{@B-}{-B@},
	moredelim=**[is][\color{blue}\bfseries]{@AB-}{-AB@},
	escapeinside={(**}{**)},
	tabsize=4,
	xleftmargin=1pt,
	captionpos=t,
	frame=bt,
	framesep=2pt,
	framerule=0.5pt,
	showstringspaces=false
}

\usepackage{float}

\definecolor{fsegreen}{rgb}{0, 0.501961, 0}
\definecolor{fsepink}{rgb}{0.90, 0.45, 0.55}

\setcopyright{cc}
\setcctype{by}
\acmDOI{10.1145/3808217}
\acmYear{2026}
\acmJournal{PACMSE}
\acmVolume{3}
\acmNumber{FSE}
\acmArticle{FSE210}
\acmMonth{7}
\acmSubmissionID{fse26mainb-p3575-p}
\received{2026-02-25}
\received[accepted]{2026-03-24}

\begin{document}

\title{Uncovering Similar but Different Packages in PyPI and Potential Security Threats}

\author{Sunha Park}
\orcid{0009-0008-1729-0696}
\affiliation{%
  \institution{Korea University}
  \city{Seoul}
  \country{Republic of Korea}
}
\email{sunhap@korea.ac.kr}

\author{Soojin Han}
\orcid{0009-0004-6734-7516}
\affiliation{%
  \institution{Dongduk Women's University}
  \city{Seoul}
  \country{Republic of Korea}
}
\email{hsjsstn@gmail.com}

\author{Seunghoon Woo}
\orcid{0000-0002-5455-0804}
\affiliation{%
  \institution{Korea University}
  \city{Seoul}
  \country{Republic of Korea}
}
\email{seunghoonwoo@korea.ac.kr}

\begin{abstract}


In this study, we present a large-scale, in-depth study of package replication in PyPI. 
As a vital platform, PyPI streamlines Python package distribution for developers.
However, beyond small-scale code cloning, we observe that many replicated packages exist on PyPI, which duplicate most of the codebase from existing packages. Such replication not only confuses developers but also propagates known vulnerabilities and enables the creation of new malicious packages. To address this issue, we comprehensively examine the characteristics and potential threats of replicated packages.
Using one-third of the entire PyPI repository (200K packages), we investigate replication from three perspectives: replication of popular packages, vulnerable packages, and malicious packages. 
Our experiments reveal three critical findings about package replication in PyPI:
(1) by identifying {1,361} replicated packages of the top 3K popular projects, we show that replication frequently redistributes substantial portions of existing packages under different maintainers;
(2) by uncovering {256} previously unknown replicated vulnerable packages, we demonstrate that replication creates vulnerability blind spots that current detection tools rarely catch;
(3) by analyzing 3,883 known malicious packages, we found that {186 (4.79\%)} replicated popular ones, and this pattern further led us to identify seven previously unknown replicated malicious packages, highlighting its role as an attack vector for malware distribution through minor modifications and code injection.
\end{abstract}

\ccsdesc[500]{Security and privacy~Software security engineering}
\ccsdesc[500]{General and reference~Empirical studies}

\keywords{Python package replication, Vulnerability propagation, Malware detection.}

\maketitle

\section{Introduction}
PyPI has grown rapidly as the central platform for distributing and reusing Python packages, hosting hundreds of thousands of projects widely used by developers worldwide~\cite{bommarito2019empirical}.
Although the platform encourages code reuse and modular development, not all reuse patterns are safe~\cite{kim2017vuddy, woo2022movery}. Beyond simple code cloning, where developers copy parts of existing projects, there is also \textit{package replication}, where most of a package's codebase is duplicated and uploaded as a new package.

{From an engineering perspective, 
this can provide high efficiency by reducing development time and costs, but at the same time, package replication may introduce security risks}~\cite{haefliger2008code}.
Malicious package replication can be exploited as the entry point for attacks such as typosquatting, which is already widely recognized as posing significant risks by misleading developers~\cite{ladisa2022taxonomy, neupane2023beyond}.
Moreover, even without malicious intent, packages replicated from those containing vulnerabilities can unintentionally propagate those security issues further~\cite{cadariu2015tracking, reid2023large, vu2020typosquatting}.

Given these concerns,
prior work has examined the security of PyPI.
Research on malicious package detection has ranged from rule-based~\cite{maloss, malwukong} to machine learning–based approaches~\cite{mphunter, ea4mp, cerebro, malguard}, and the construction of public malicious package datasets~\cite{ohm2020backstabber, guo2023empirical}.
Other studies have investigated vulnerabilities and maintenance issues in Python packages~\cite{alfadel2023empirical, valiev2018ecosystem}. 
However, their goal is to identify insecure packages in PyPI, which is rather distant from detecting replicated packages or analyzing replication to uncover insecure packages. 
Beyond PyPI, prior work in other ecosystems has explored code clones~\cite{jiang2007deckard, jang2012redebug, sajnani2016sourcerercc, li2016vulpecker, woo2022movery}, orphan vulnerabilities~\cite{reid2022extent, reid2023large}, shrinkwrapped clones in npm~\cite{wyss2022fork}, and large-scale duplication on GitHub~\cite{lopes2017dejavu}. Nevertheless, no large-scale study has systematically examined package replication in the Python ecosystem from a security perspective.
%



In this study, we aim to closely analyze this problem and address the following research questions.

\vspace{-0.3em}
\begin{itemize}
    \setlength\itemsep{0.1em}
    \item[\textbf{RQ1.}] \textbf{Popularity analysis.} How prevalent are replicated packages in PyPI, and what are their fundamental characteristics (\eg, naming and code similarity)? 

    \item[\textbf{RQ2.}] \textbf{Vulnerability analysis.} How widespread are vulnerabilities in replicated packages, and what are their primary characteristics (\eg, CWE categories and CVSS scores)? 

    \item[\textbf{RQ3.}] \textbf{Malware analysis.} What are the characteristics of replicated malicious packages? 
    Can these characteristics help identify unknown replicated malicious packages?
\end{itemize}
\vspace{-0.3em}


{Unlike traditional code clones that involve reusing only small fragments of code, we define a \textit{replicated package} as a new package that is created by reusing a substantial portion of an existing package's codebase (\eg, two packages are considered highly replicated when 
their code similarity exceeds 90\%; see \mbox{\autoref{subsec:res}}).}
{Although prior research has discussed partial code reuse, the identification and characterization of replicated packages and their security implications have not been thoroughly examined. We address these gaps by answering the three key research questions.}

To this end, we constructed five datasets: (1) \textit{popularity}, (2) \textit{vulnerability}, (3) \textit{malware}, (4) \textit{recent}, and (5) \textit{candidate} datasets (as of Aug. 2025). The candidate dataset includes the latest versions of approximately one-third of all PyPI packages (200K out of 670K) and is used to analyze replication trends and identify vulnerable packages. The popularity dataset covers the top 3K packages by downloads, while the vulnerability and malware datasets contain known vulnerable and malicious packages. The recent dataset was collected to evaluate the security of newly published packages (see \autoref{subsec:dataset}).
To identify replicated packages, we propose a framework combining embedding, clustering, and code similarity analysis. 
Packages are embedded using \texttt{CodeT5+}~\cite{wang2023codet5plus} (see \autoref{subsec:embed}), followed by
dimensionality reduction with \texttt{UMAP}~\cite{mcinnes2018umap} and density-based clustering via \texttt{HDBSCAN}~\cite{mcinnes2017hdbscan}. Replication is then confirmed through fine-grained code similarity analysis (see \autoref{subsec:res}).


Our experiments reveal three critical findings: (1) \textit{replication frequently redistributes substantial portions of existing packages}, often under different maintainers, (2) \textit{replication creates vulnerability blind spots} that are difficult for current tools to detect, and (3) \textit{replication provides an attack vector for malware distribution} through minimal modifications combined with malicious code injection.

From the candidate dataset, we identified {1,361} replicated packages derived from popular projects, {60\%} of which were maintained by different authors than the originals (see \autoref{sec:rq1}).
To evaluate security implications, we mapped 1,072 known vulnerable packages to the candidate dataset and discovered {256} previously unknown replicated vulnerable packages; {35.14\%} were high or critical severity and had remained unmaintained for an average of {1,600} days.
Existing dependency scanners missed over 98\% of these cases, exposing a critical blind spot in current vulnerability management approaches that rely on package names and versions (see \autoref{sec:rq2}).
%
We further analyzed malicious packages and found that, among 3,883 known malicious packages, {186} were replication-based variants with injected suspicious APIs (\eg, \texttt{arbitrary} \texttt{code} \texttt{execution}). Applying our method to newly registered PyPI packages uncovered seven previously unknown malicious packages, all later removed after disclosure (see \autoref{sec:rq3}).

\PP{Contribution} This paper makes the following four main contributions.

\begin{itemize}
    \setlength\itemsep{0.3em}
    \item \textit{Large-scale analysis.} To analyze overall insights and potential risks of package replication, we constructed five datasets and conducted large-scale experiments. In particular, the candidate dataset was built to represent PyPI, consisting of the latest versions of 200,737 repositories.
    \item \textit{Thorough study.}
    We conducted a thorough and multi-faceted analysis of package replication. Our experiments did not rely solely on name or metadata similarity; instead, we incorporated clustering and code similarity techniques, and designed specialized experiments for vulnerability and malware detection to ensure a more rigorous evaluation.    
    \item \textit{Practicality.}
    Our experiments showed a practical impact by reporting and removing high-risk vulnerable and malicious packages. We also exposed the limits of existing techniques and proposed a replication-based method as a practical way to strengthen PyPI's security.
    \item \textit{Open source.} 
    Our source code and experimental results are available at: \url{https://github.com/sunha21/pypi-replication-analysis}.
\end{itemize}
\section{Methodology}\label{sec:metho}

\begin{figure}[t]
	\begin{center}
		\includegraphics[width=0.95\linewidth]{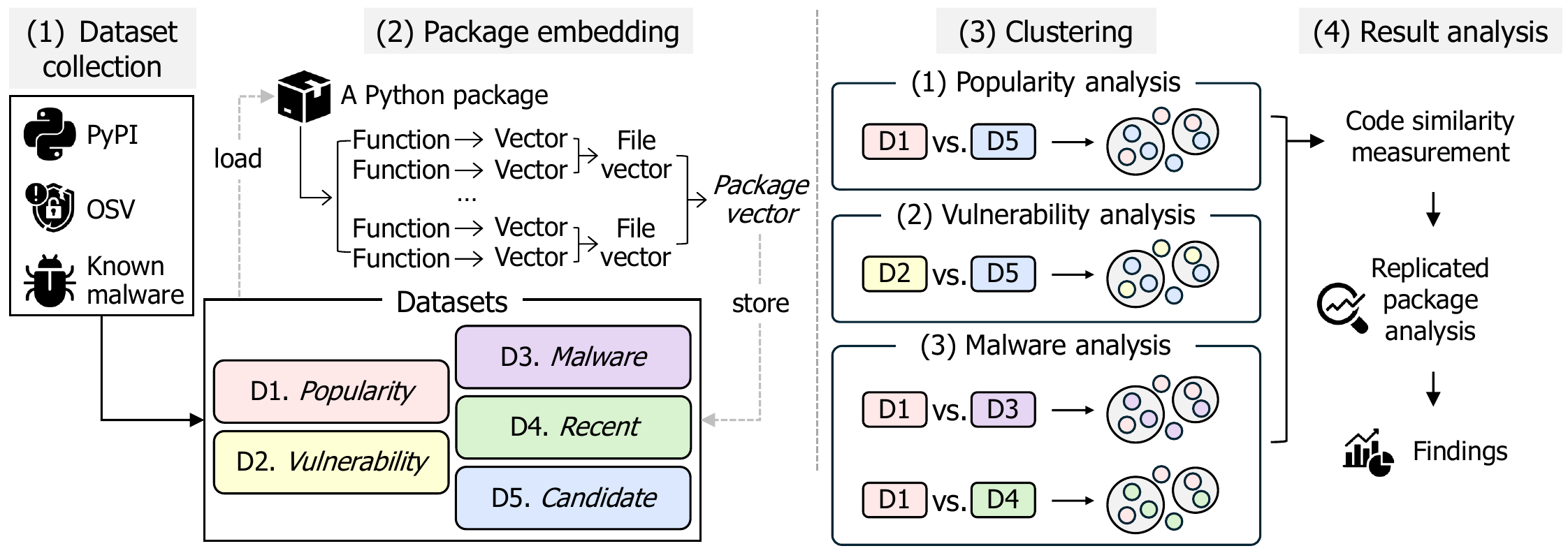}	
		
		\vspace{-0.5em}
		\caption{\label{fig:overview}Overview of the empirical study.} 
	\end{center}
	
	\vspace{-1.0em}
\end{figure}

\autoref{fig:overview} illustrates the high-level workflow of this study. To comprehensively analyze package replication, we constructed five datasets and embedded all packages to generate package-level vectors. These vectors were then clustered according to the analysis objectives, followed by additional code similarity analysis and other inspections to perform a deep analysis of package replication.





\subsection{Dataset Collection}\label{subsec:dataset}

We construct the following five datasets: (1) \textit{popularity}, (2) \textit{vulnerability},
(3) \textit{malware},
(4) \textit{recent}, 
and (5) \textit{candidate} datasets. \autoref{table:dataset} provides a quantitative overview of the collected datasets.

\subsubsection{Popularity Dataset}
The popularity dataset comprises the most popular packages in PyPI and becomes the reference point for identifying replications.
To construct this dataset, we collected the codebases of the top 3,000 packages from PyPI (as of August 2025) based on the Top PyPI Packages dataset~\cite{hugovk2025}. Each of these packages had over one million downloads in August 2025, making them a reasonable standard for popular packages. We used the official PyPI JSON API to download all versions of the packages and extracted only the Python source files.  
While we initially targeted 3,000 packages, some contained no Python source or no valid functions. After filtering, the dataset comprises 2,767 packages, 160,317 versions, and 8,701,101,354 lines of code. Note that all dataset statistics reported in this section are measured after filtering out non-Python and functionless files.

\subsubsection{Vulnerability Dataset}

We classified any package with one or more vulnerabilities reported in PYSEC (Python Security Database) and GHSA (GitHub Security Advisory) from Google's Open Source Vulnerabilities (OSV)~\cite{osv} as a vulnerable package. 
%
Consequently, we collected all versions of the 1,072 identified vulnerable packages, resulting in 70,280 package versions with 4,628,177,354 accumulated lines of code for analysis.
This set includes not only vulnerabilities related to input validation but also various types, such as path traversal and cross-site scripting.


\subsubsection{Malware Dataset}
The malicious dataset was collected from publicly available sources: Guo \etal~\cite{guo2023empirical} and Ohm \etal~\cite{ohm2020backstabber}.
They include various types of malicious Python packages (\eg, download-driven, typosquatting, and dependency confusion).
As a result, we obtained 2,656 malicious packages comprising 3,883 versions (with 7,291,220 accumulated lines of code), which are used to examine the prevalence and characteristics of duplicated malicious code.

\begin{table}[t]
\centering
\renewcommand{\tabcolsep}{3mm}	
	\caption{\label{table:dataset}Overview of the five datasets collected for experiments.}
    \vspace{-1em}
    \small
	\begin{center}	
\begin{tabular}{|c|r|r|r|r|r|}
\hline
\multicolumn{1}{|c|}{\rule{0in}{2.2ex}\textbf{Dataset}}
& \multicolumn{1}{c|}{Popularity}
& \multicolumn{1}{c|}{Vulnerability}
& \multicolumn{1}{c|}{Malware}
& \multicolumn{1}{c|}{Recent}
& \multicolumn{1}{c|}{Candidate}\\\hline\hline
%
\rule{0in}{2.2ex}\textbf{\#Packages}
& 2,767
& 1,072
& 2,656
& 25,407 
& 200,737 
\\\hline
\rule{0in}{2.2ex}\textbf{\#Versions}           
& 160,317
& 70,280          
& 3,883                             
& 36,008 
& 200,737
\\\hline
\rule{0in}{2.2ex}\textbf{\#Lines of Code}      
& 8,701,101,354                                
& 4,628,177,354
& 7,291,220
& 196,760,716 
& 1,020,502,964
\\\hline
\end{tabular}
\end{center}
\vspace{-1.5em}
\end{table}

\subsubsection{Recent Dataset}
This dataset is constructed to capture the trends of the most recent PyPI packages.
From May to August 2025, we collected newly registered packages by monitoring the PyPI RSS feed.  
For each package, we downloaded all available versions at the time of discovery, resulting in 25,407 packages and 36,008 versions with 196,760,716 accumulated lines of code.

\subsubsection{Candidate Dataset}
Finally, the candidate dataset consists of general packages from PyPI and is regarded as the pool for replication detection.
However, collecting all packages from PyPI is resource-inefficient and imposes excessive load during analysis. Therefore, we randomly sampled a subset of packages, representing approximately one-third of the 670,000 packages available on PyPI (as of August 2025), and collected the codebases of their latest versions to construct the candidate dataset. 
In total, the dataset comprises 200,737 packages and more than 1B lines of Python code.


\subsection{Package Embedding}\label{subsec:embed}
To address our research questions, we perform comparisons across the constructed datasets.
For example, to address RQ1 (\ie, popularity analysis), we need to compare the packages in the popularity dataset with those in the candidate dataset.
However, direct comparison of the raw codebase is inefficient due to its inherent code diversity, as code can be modified during package replication.

To identify replicated packages, therefore, 
we embed the packages belonging to each dataset and then apply clustering across embeddings from different datasets (see \autoref{subsec:clustering}).
Specifically, we embed all the packages collected in the dataset into 256-dimensional vectors.

Before embedding packages, we preprocess the source code by extracting functions and removing all comments to focus solely on functionality and semantics. For function extraction, \texttt{ctags}~\cite{ctags} was adopted due to its lightweight and efficient parsing, which scales well to large codebases.

For package embedding, we employ \texttt{CodeT5+}~\cite{wang2023codet5plus}, a transformer-based large language model designed for code understanding and generation.
Its strong performance on clone detection makes it particularly suitable for identifying semantic similarities in source code~\cite{niu2023empirical}. 
To maintain code semantics and ensure consistent analysis, 
we first apply \texttt{CodeT5+} at the function level.
Let $P$ be a Python package represented as a set of files, 
and let each file $F_i$ be represented as a set of functions.

\vspace{-0.8em}
\[
P = \{ F_1, F_2, \ldots, F_m \}, \quad
F_i = \{ f_{i1}, f_{i2}, \ldots, f_{in_i} \}.
\]
\vspace{-1.0em}

Let $\phi$ be the embedding function implemented using \texttt{CodeT5+}, 
which maps a function $f_{ij}$ into a 256-dimensional vector.

\vspace{-1.3em}
\[
v_{ij} = \phi(f_{ij}) \in \mathbb{R}^{256}
\]
\vspace{-1.0em}

Here, we applied a sliding window approach with a stride of 128 tokens to handle long functions, since the model accepts a maximum input length of 512 tokens per function and many real-world functions exceed this limit.
Each function is then encoded into a 256-dimensional vector.

Next, we obtain the file vectors by applying a mean operation over the function vectors contained in each file.
Let $v_{F_i}$ be the vector representation of file $F_i$, defined as the mean of its function vectors.

\vspace{-0.5em}
\[
v_{F_i} = \frac{1}{n_i} \sum_{j=1}^{n_i} v_{ij}
\]
\vspace{-0.4em}

Finally, we obtain the package vector by applying a mean operation over its file vectors.
Let $v_{P}$ be the vector representation of package $P$, defined as the mean of its file vectors.

\vspace{-0.5em}
\[
v_{P} = \frac{1}{m} \sum_{i=1}^{m} v_{F_i}
\]
\vspace{-0.4em}


Through this process, each package version is ultimately represented as a single 256-dimensional vector, which enables efficient similarity computation across packages.

\subsection{Clustering}\label{subsec:clustering}
In our experiments, clustering algorithms are required to assess the similarity of embedded packages. 
Clustering is performed between two different datasets depending on the research question. 
However, because the generated package vectors are 256-dimensional, performing effective clustering directly on them is challenging.
This is because high-dimensional spaces suffer from the curse of dimensionality, where distances become less meaningful and data sparsity increases, substantially degrading clustering performance.

To address this issue, we first perform dimensionality reduction. Specifically, we adopt \texttt{UMAP}~\cite{mcinnes2018umap}, which effectively maps high-dimensional code embeddings into a lower-dimensional space while preserving the underlying data structure. This property makes \texttt{UMAP} well-suited for clustering tasks on code embeddings.

For clustering, we adopt \texttt{HDBSCAN}~\cite{mcinnes2017hdbscan}, which does not require a predefined number of clusters.
In practice, the number of similar package groups is unknown, and specifying the number of clusters in advance can lead to significant detection errors. The algorithm can identify clusters of varying densities, making it appropriate for package detection where different types of packages may exhibit different clustering characteristics. 

Formally, given the set of package vectors $\mathcal{V} = \{ v_{P_1}, \ldots, v_{P_N} \}$ and the dimensionality reduction function $\psi$ (\texttt{UMAP}), 
we obtain lower-dimensional embeddings $\tilde{v}_{P_i} = \psi(v_{P_i})$. 
The clustering function $\mathcal{C}$ (\texttt{HDBSCAN}) is then applied as follows.
\[
\{ C_1, C_2, \ldots, C_K \} = \mathcal{C}\big( \psi(\mathcal{V}) \big),
\]
where each $C_k$ denotes a cluster of similar packages.

To capture small but meaningful clusters effectively while reducing outliers, we set the minimum cluster size (\texttt{min\_cluster\_size}) to 2, the minimum number of samples (\texttt{min\_samples}) to 1, and the option \texttt{cluster\_selection\_epsilon} to 0.15. These values were determined empirically through parameter tuning. In addition, clustering too many package vectors at once may reduce accuracy, thus we divided the candidate dataset into batches and clustered each batch separately.

\subsection{Identifying Replicated Packages}\label{subsec:res}


Finally, we examine the clustering results and provide insights into replicated packages. However, 
simply assuming that all packages within the same cluster constitute replications may undermine accuracy.
%
{To address this, 
within each cluster, we computed detailed code similarity scores for all possible package pairs.}


\subsubsection{Code Similarity Measurement}\label{subsubsec:code_sim}
Let the two packages under comparison be denoted as $X$ and $Y$. We compare both the file paths and the file contents of $X$ and $Y$. Each pair of files is classified into one of the following four categories: \textit{identical}, \textit{modified}, \textit{added}, or \textit{deleted}.

\begin{itemize}
    \setlength\itemsep{0.8em}
    \item \textit{Identical.} The file in $X$ and the file in $Y$ share the same path (including the immediate parent directory name) and have exactly the same file contents (\ie, code syntax).

    \item \textit{Modified.} The file paths match, but the file contents differ.

    \item \textit{Added.} A file exists in $Y$ such that no file in $X$ has both the same path and identical contents.

    \item \textit{Deleted.} A file exists in $X$ such that no file in $Y$ has both the same path and identical contents.
\end{itemize}

Based on these classifications, we computed a code similarity score between the two packages.

\vspace{0em}
\[
\text{\texttt{sim}}(X,Y) = 
\frac{\#\text{identical} + (0.5 \times \#\text{modified})}
{\#\text{identical} + \#\text{modified} + \#\text{added} + \#\text{deleted}}
\]
\vspace{0em}

The weight of 0.5 was empirically determined from our candidate dataset. We first defined a preliminary similarity score based on file identity: the number of identical files divided by the total number of files in the union of the original and replicated packages.
Among the 2,350 packages with a similarity score of at least 0.1 and at least one modified file, the average line-level modification ratio was 0.44. Based on this observation, we assign a weight of 0.5 to modified files.

\subsubsection{Definition of Package Replication}
Given two packages $X$ (reference package) and $Y$ that belong to the same cluster, we define the replication level of $Y$ relative to $X$ based on their similarity score \texttt{sim}($X, Y$) as follows.
\vspace{0.4em}
\begin{itemize}
    \setlength\itemsep{1.0em}
    \item If \textbf{\texttt{sim}($\boldsymbol X$, $\boldsymbol Y$) $\boldsymbol\geq$ 0.9}, $Y$ is considered a \textbf{highly replicated package} of $X$.
    
    \item If \textbf{0.5 $\boldsymbol\leq$ \texttt{sim}($\boldsymbol X$, $\boldsymbol Y$) $\boldsymbol<$ 0.9}, $Y$ is considered a \textbf{partially replicated package} of $X$.
\end{itemize}
\vspace{0.4em}

%
In our experiments, all replicated packages underwent additional inspection. {Although no strict guidelines exist, we adopted 0.9 as a practical boundary for near-identical packages and 0.5 as the cutoff for partial similarity.
\mbox{\autoref{subsec:simsen}} presents the sensitivity analysis of this threshold.}

\subsubsection{Name and Metadata Similarity}

Package names and metadata fields are the primary sources of information users rely on to identify and install Python packages. 
To measure name similarity, we computed the \texttt{Levenshtein} ratio between two package names. This ratio normalizes edit operations by the combined length of the two names, where a score of 1 indicates identical names.

For metadata, we considered five fields (\texttt{author}, \texttt{author\_email}, \texttt{summary}, \texttt{description}, and \texttt{home\_page}), and performed exact string matching for each field.
%

\subsubsection{Human Resources}
{To answer the research questions, the identified replicated packages were analyzed by three experts.}
One has more than ten years of experience in software engineering and security, while the other two have three and five years of experience, respectively. The manual analysis mainly involves directly reviewing source code or comparing metadata, and, when necessary, includes triggering vulnerabilities or testing malicious code in a sandbox environment.

\section{Evaluation}
{We first evaluate our replication detection approach to validate its effectiveness and to enhance the credibility of our findings in addressing the research questions.}

\subsection{Replicated Package Detection Accuracy}\label{subsubsec:acc}

\subsubsection{Methodology}
{To assess accuracy, we construct a validation set and adopt a comparative validation strategy using a code clone detection tool as a reference.
This is necessary because no ground-truth dataset exists for package-level replication, and defining replication at this level is inherently challenging: metadata-based signals are insufficient, and code-level similarity often requires subjective judgment.}
{Therefore, we selected SourcererCC~\mbox{\cite{sajnani2016sourcerercc}}, a scalable, language-agnostic code clone detection tool based on token similarity. 
Because SourcererCC operates at the file level (\ie, detecting similar files using token-level similarity), we extend its output to compute package-level similarity. For a package pair $X$ and $Y$, we measure the proportion of files involved in clone relationships across both packages using an 80\% token similarity threshold (default):}

\vspace{-1.7em}
\[
\text{\texttt{sim}}_{SCC}(X,Y) =
\frac{|F_X^{clone}| + |F_Y^{clone}|}
{|F_X| + |F_Y|}
\]
\vspace{-1.3em}

where $F_X$ and $F_Y$ denote the sets of files in packages $X$ and $Y$, while $F_X^{clone}$ and $F_Y^{clone}$ represent the files that participate in at least one clone relationship detected by SourcererCC.

{We use the popularity dataset (2,767 packages) as the baseline and compare it against the candidate dataset (200,737 packages). Because SourcererCC is not designed to operate at this scale, we adopt a filtered evaluation strategy. Specifically, we first apply our replication detection approach and retain package pairs with a similarity score above 0.1, excluding clearly unrelated pairs while preserving borderline cases. This results in 3,086 candidate pairs, which are analyzed using SourcererCC.}

\subsubsection{Criteria for Result Analysis}
{The 3,086 candidate pairs were manually analyzed and classified into replicated and non-replicated packages. To reduce ambiguity, we first filtered out clear-cut cases: pairs with similarity scores $\geq$ 0.7 in both tools were labeled as replicated, whereas those with similarity scores $<$ 0.3 in both tools were labeled as non-replicated.}

{For the remaining ambiguous cases, we prioritized labeling according to the following criteria. First, we classified a pair as replication when it exhibited strong replication signals, \ie, high name similarity ($\geq$ 0.7) together with substantial structural similarity ($\geq$ 0.5 in at least one tool or $\geq$ 0.3 in both tools).}
{Next, for cases involving vendoring, we labeled a pair as replication only when one package was predominantly derived from the other and further extended or modified it. If the similarity arose because one package included the other as just one component among multiple incorporated libraries, we treated the relationship as ordinary code reuse (\ie, non-replication).}
{Finally, the detected package pair was not considered replicated when the observed similarity was limited to small fragments or boilerplate code.}





{Any cases that could not be conclusively determined under these criteria were finally resolved based on agreement among the participating analysts. 
As a result, from the 3,086 candidate pairs, we identified 1,767 replicated pairs and 1,319 non-replicated package relationships.}

\subsubsection{Result Analysis}
{\mbox{\autoref{table:acc_res}} presents the comparison between our approach and SourcererCC under a replication threshold of 0.5 for both methods. Although the scoring methods differ, this threshold indicates cases where at least half of the package-level content is considered replicated.}

{Overall, because SourcererCC performs finer-grained clone analysis, it achieved a slightly higher recall (78.61\% vs. 77.02\%). However, our approach yielded substantially higher precision (91.58\% vs. 82.04\%), resulting in a superior F1-score and overall accuracy.}

{Notably, SourcererCC generated numerous FPs for small packages. Incidental code overlap cases were frequently assigned high similarity scores and incorrectly labeled as replication. Moreover, similarity driven by commonly included boilerplate files, such as \texttt{setup.py}, further contributed to FP classifications. In addition, when code modifications were introduced during the package replication process and sufficient file-level clones were not detected, SourcererCC reported FNs.}

{Our approach incorporates file names as well as added and deleted files into the similarity computation, which makes it more robust to incidental code overlap cases and results in fewer FPs. However, our approach produced several FNs. Because modified files are assigned uniform weight, near-identical but slightly altered files fall below the replication threshold (\ie, FNs). Furthermore, replication involving extensive file additions or deletions reduces similarity and escapes detection.}

{Despite the identified FPs and FNs, the achieved accuracy provides a reliable foundation for analyzing the validated true positive replication cases in the subsequent RQs.}


\begin{table}[t]
\centering
\renewcommand{\tabcolsep}{2.5mm}	
\caption{\label{table:acc_res}{Results of replicated package detection accuracy evaluation (TP: True Positive, FP: False Positive, FN: False Negative, TN: True Negative).}}
\vspace{-1em}
\small	
\begin{tabular}{|c||c|c|c|c||c|c|c|c|}
\hline
\rule{0in}{2.2ex}\textbf{Metric}
& \#TP
& \#FP
& \#FN
& \#TN
& Precision
& Recall
& F1-score
& Accuracy\\\hline\hline
\rule{0in}{2.2ex}\textbf{Our approach}
& 1,361 & 125 & 406 & 1,194 & \textbf{91.58\%} & 77.02\% &\textbf{83.68\%} & \textbf{82.79\%}\\\hline
\rule{0in}{2.2ex}\textbf{SourcererCC}~\cite{sajnani2016sourcerercc}
& 1,389 & 304 & 378 & 1,015 & 82.04\% & \textbf{78.61\%} & 80.29\% & 77.90\%\\\hline
\end{tabular}
\vspace{-1.5em}
\end{table}




\subsubsection{Threats to Validity}
{To mitigate subjectivity, we labeled ambiguous cases using similarity thresholds and structural criteria. Nonetheless, labeling bias and human error may occur. Moreover, our extension of SourcererCC to package-level similarity has not been independently validated; thus, the comparison should be interpreted as relative rather than absolute. Finally, because SourcererCC frequently maps a single file to multiple files across packages, it tends to inflate similarity scores, potentially yielding higher FPs and lower FNs than its intrinsic performance would indicate.}

\subsubsection{Performance}
{In the experiment comparing the popularity dataset (2,767 packages) with the candidate dataset (200,737 packages), we measured the execution time of the three stages in our pipeline: embedding, clustering, and similarity computation.}
{All experiments were conducted on a server equipped with two NVIDIA RTX A6000 GPUs (48GB GDDR6 each), an AMD Ryzen Threadripper Pro 7965WX processor (24 cores, 48 threads), 384GB ECC RAM, and 4TB SSD.}

{For embedding, each package version required less than five seconds on average. Clustering across 200 batches required a total of 1,860 s. 
Finally, within each cluster, we first identified candidate similar package pairs and then computed their similarity scores. Processing approximately 2.74 million such pairs required 36 hours in total, corresponding to about one second per pair on average.}
%
{Our choice of file-level granularity, a relatively coarse abstraction, combined with efficient and scalable techniques and encoding strategies, results in a replication detection algorithm that is inherently scalable. 
In contrast, although SourcererCC was applied only to the specific software version pairs identified after our clustering stage and performed similarity computation, it still required five seconds per package pair on average.
In practice, the process must also consider identifying the most similar version pairs within each cluster. Therefore, this difference further underscores that our approach is designed with scalability as a primary consideration.}

\begin{table}[t]
\centering
\renewcommand{\tabcolsep}{2mm}	
\caption{\label{table:clustering_p}{Sensitivity analysis of clustering parameters.}}
\vspace{-1em}
\small
\begin{tabular}{|c|c|c|c|c|c|}
\hline
\rule{0in}{2.2ex}\textbf{\texttt{min\_cluster\_size}}
& \textbf{\texttt{min\_samples}}
& \textbf{\texttt{epsilon}}
& \textbf{Avg. \#Clusters}
& \textbf{Avg. \#Outliers}
& \textbf{\#Pairs ($\geq$ 0.5)}\\\hline\hline
\rule{0in}{2.2ex}5 & 20 & 0    & 207  & 5588 & 54 \\
5 & 5  & 0    & 908  & 2789 & 76 \\
5 & 5  & 0.15 & 728  & 1907 & 73 \\
3 & 2  & 0.15 & 1000 & 746  & 84 \\
2 & 2  & 0    & 1960 & 2439 & 71 \\
\textbf{2} & \textbf{1} & \textbf{0.15} & \textbf{1232} & \textbf{393} & \textbf{94} \\\hline
\end{tabular}
\vspace{-1.5em}
\end{table}

\subsection{Clustering Threshold Sensitivity}
{We then examined how the parameters of HDBSCAN affect cluster structure. In this study, clustering is employed as a preprocessing step to generate candidate pairs across the entire package space, rather than the final stage for determining replication packages. Therefore, we adjusted the parameters to maximize the range of candidates searched for.} 

{To strengthen the observed trends without evaluating the full dataset, we conducted sensitivity analysis on randomly selected 20 (out of 200) batches. 
\mbox{\autoref{table:clustering_p}} presents the results of the clustering threshold sensitivity analysis. 
Increasing \texttt{min\_cluster\_size} filtered out small clusters, while increasing \texttt{min\_samples} classified more boundary points as noise. The \texttt{epsilon} parameter controlled the merging of nearby dense regions. Overall, relaxing density constraints reduced outliers and increased candidate clusters, thereby expanding the pool of potentially replicated packages.}


{Therefore, to minimize outliers and reduce excessive splitting, we set \texttt{min\_cluster\_size} to 2, \texttt{min\_samples} to 1, and \texttt{epsilon} to 0.15. Using this configuration, clustering produced an average of 1,232 clusters and 393 outliers per batch (across 200 batches in total).}







\section{Popularity Analysis (RQ1)}\label{sec:rq1}
We then aim to answer RQ1 by analyzing popular packages on PyPI to determine how extensively they have been replicated, as well as the causes and characteristics. {To this end, we provide a detailed analysis of the results obtained from accuracy evaluation (\mbox{\autoref{subsubsec:acc}}).
This study does not aim to analyze the complete set of ground-truth replications. Instead, we focus on the characteristics of replications identified by our similarity metric and categorized as high or partial similarity. Replications with similarity scores below 0.5 were excluded, as including them would undermine analytical consistency and reflect fragment-level code cloning rather than package-level replication.}



\subsection{Status of Popular Package Replications}

\subsubsection{Status}
We examine how extensively 2,767 popular packages are replicated among 200,737 PyPI packages. Consequently, we identified {\textbf{1,361}} replicated packages: {\textit{334 (24.54\%) highly}} replicated packages and {\textit{1,027 (75.45\%) partially}} replicated packages.
This observation suggests that replication manifests in different forms, from near-complete copies to partial overlaps.
Considering the current size of PyPI ($\approx$ 670,000 packages as of August 2025), we estimate that more than {4,000 packages ($\approx$ 1,361 $\times$ 3)} in the entire ecosystem may exhibit duplication behavior.

Although the proportion of replicated packages appears relatively small at {0.6\%}, this result is still meaningful. Given the vast scale of the PyPI ecosystem, even a fraction of a percent amounts to more than a thousand replicated packages. This scale indicates that replication is not a negligible phenomenon and should be considered when analyzing ecosystem quality.

\vspace{-0.3em}
\begin{tcolorbox}[colback=lime!13, 
    colframe=black, 
    boxrule=0pt, 
    left=5pt,
    right=5pt,
    top=2pt,
    bottom=2pt,
    enhanced jigsaw, 
    sharp corners]
    \textit{\textbf{Finding 1.} We identified {1,361} replicated packages, many of which reproduce the majority of the original package's code.
    This highlights that replication often involves extensive reuse of the original implementation, making it a non-negligible phenomenon affecting ecosystem quality.}
\end{tcolorbox}
\vspace{-0.5em}

\subsubsection{Causes of Replication}
Through manual inspection, we inferred the reasons for package replication. 
The most common reason is \textit{customization}, where developers copy packages to extend or modify their functionality. 
Among these cases, many involved only minor changes, such as minimal edits to the \texttt{setup.py} file.
In fact, some packages reproduce the original codebase exactly without any modifications, which appears to be intended for mirroring or rehosting.

Another frequent purpose is vendoring (or internal forking), where external dependencies are cloned to operate within isolated environments, reduce dependency risks, or apply controlled updates. These replications often preserve the exact code of the original package but are redistributed under a different name.
Some replications were created for backward compatibility, such as maintaining support for Python 2. In these cases, developers re-released the same package with only minimal changes to ensure compatibility with legacy environments.
Finally, we observed some replications with malicious intent. In such cases, attackers insert malicious code into the replicated package and redistribute it, thereby exploiting replication as a way for malware propagation. This will be examined in detail in the experiments for RQ2 and RQ3.
\vspace{-0.3em}

\begin{tcolorbox}[colback=lime!13, 
    colframe=black, 
    boxrule=0pt, 
    left=5pt,
    right=5pt,
    top=2pt,
    bottom=2pt,
    enhanced jigsaw, 
    sharp corners]
    \textit{\textbf{Finding 2.} In many cases, package replication was driven by specific customization needs. 
    In some cases, it was exploited for malicious purposes. 
    {These observations suggest that large-scale replication warrants ecosystem analysis to better understand its maintenance and security implications.}
    }
    %
\end{tcolorbox}
\vspace{-0.5em}

\subsection{Characteristics of Replicated Packages}


\subsubsection{Maintainer}\label{subsubsec:rq1_maintainer}
We first analyzed the ownership of packages.
Because the \textit{author} field can be arbitrarily set, 
we focus on the \textit{maintainer} displayed on the PyPI website. 
Of the {1,361} detected replicated packages, {817 (60.03\%)} were distributed by accounts different from those of the original packages, indicating that the vast majority of replications were published by unrelated maintainers.
%

{When replicated packages are maintained by the same maintainer, they may reflect legitimate scenarios such as renaming or restructuring. Replication by different maintainers, however, introduces a governance boundary between the original and the replicated package. Although code reuse and modification are fundamental principles of open-source ecosystems, near-full replication under separate maintainership can hinder patch propagation, obscure accountability, and fragment maintenance efforts. In such cases, security fixes applied upstream may not be consistently adopted downstream, potentially leaving replicated variants exposed.}

\vspace{-0.3em}
\begin{tcolorbox}[colback=lime!13, 
    colframe=black, 
    boxrule=0pt, 
    left=5pt,
    right=5pt,
    top=2pt,
    bottom=2pt,
    enhanced jigsaw, 
    sharp corners]
    \textit{\textbf{Finding 3.} {39.97\% of replicated packages were created by the same maintainer, whereas 60.03\% were maintained by different maintainers. Although replication is not inherently problematic, the high proportion of cross-maintainer replication raises concerns such as patch consistency.}
    }
    %
\end{tcolorbox}
\vspace{-0.5em}


\subsubsection{Name and Metadata Similarity}
We classify replicated packages by whether they share the same maintainer ({544}) or not ({817}) and compare their names and metadata with the originals to distinguish legitimate republishing from potentially malicious or unauthorized replication.


\begin{figure}[t]
	\begin{center}
         \begin{subfigure}[b]{0.44\textwidth}
			\centering
                \includegraphics[width=0.85\linewidth]{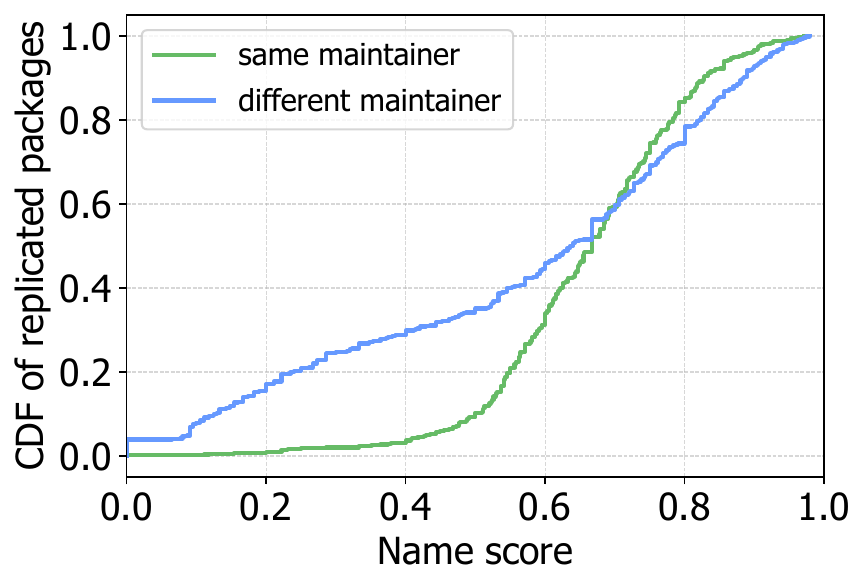}
			
			\vspace{-0.3em}
			\caption{Distribution of \textbf{name similarities} between original and replicated packages.}
			\label{fig:name-cdf}
            \vspace{0.5em}
		\end{subfigure}\hspace{1em}%
		\begin{subfigure}[b]{0.44\textwidth}
			\centering
                \includegraphics[width=0.85\linewidth]{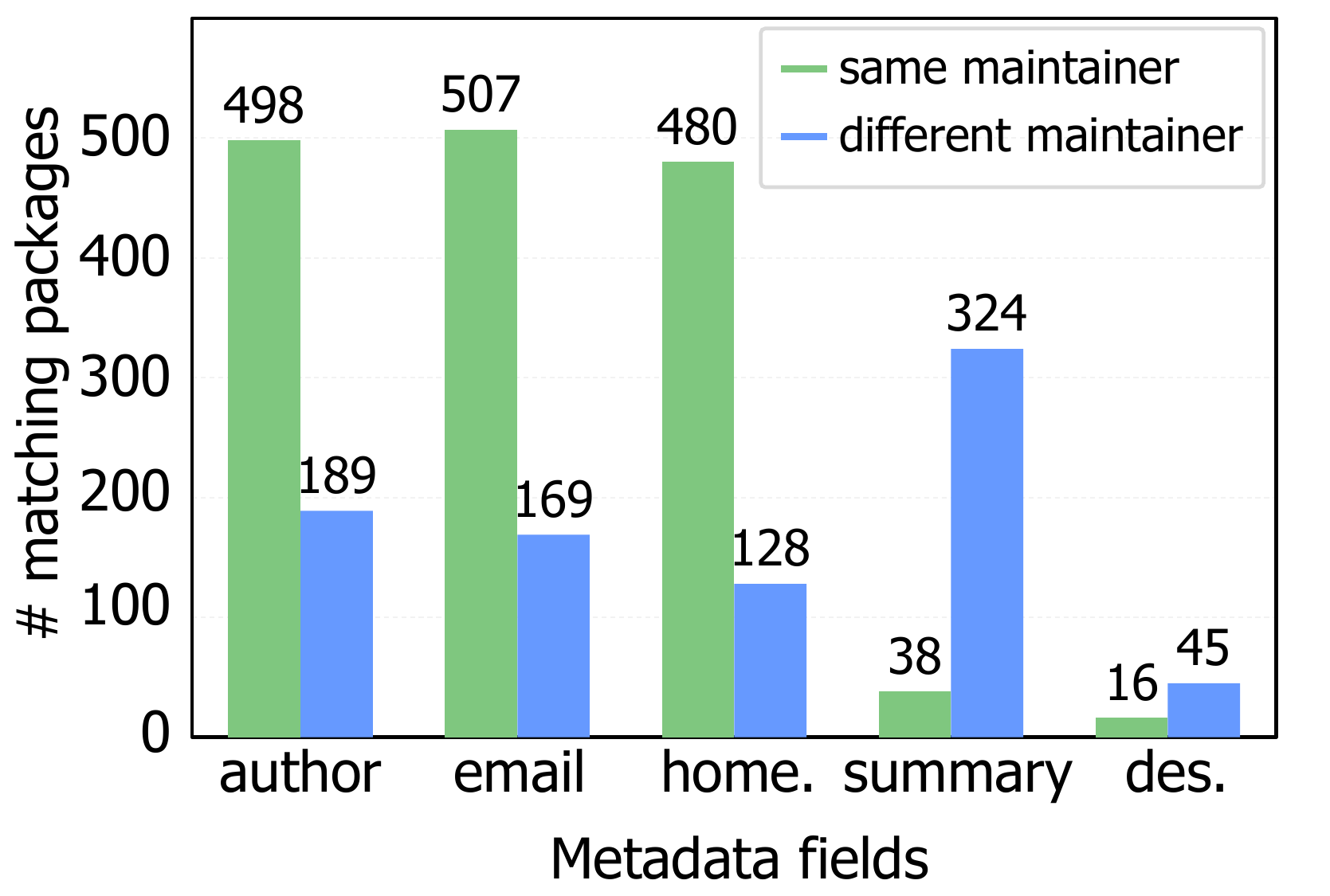}
			
			\vspace{-0.3em}
			\caption{Illustration of \textbf{metadata} matches (original vs. replicated).}
			\label{fig:meta}
            \vspace{0.5em}
		\end{subfigure}
        \vspace{-0.5em}
		\caption{Graphs on name and metadata similarity. Packages maintained by the same maintainer generally have more similar names and a higher possibility of metadata matching.}\label{fig:theta}
	\end{center}
	
	\vspace{-1.0em}
\end{figure}

\PP{Replication by the Same Maintainer (544 Packages)}
%
\autoref{fig:name-cdf} presents the cumulative distribution function (CDF) of name similarity.
For replicated packages maintained by the same maintainer, 
the average name similarity was {$0.65$ (median $0.72$)}.
Notably, {222} of {544} packages exhibited high similarity ($\geq$ 0.7), indicating that maintainers often reused the original name with minor modifications, such as adding a prefix or suffix (\eg, \texttt{Keras} (original) and \texttt{Keras\_nightly} (replicated)).
For metadata similarity, {491} packages ({89.4\%}) retained at least three identical metadata fields.
As shown in \autoref{fig:meta}, the fields that were most frequently matched were \texttt{author\_email} ({507}), followed by \texttt{author} ({498}) and \texttt{home\_page} ({480}).
When metadata differed, packages often shared similar code but served different functional purposes (\eg, \texttt{safetycli} and \texttt{safety}).


\vspace{-0.3em}
\begin{tcolorbox}[colback=lime!13, 
    colframe=black, 
    boxrule=0pt, 
    left=5pt,
    right=5pt,
    top=2pt,
    bottom=2pt,
    enhanced jigsaw, 
    sharp corners]
    \textit{\textbf{Finding 4.} 
    For replicated packages with the same maintainer, {222} out of {544} packages had highly similar names to their original counterparts. However, the summary and description fields often differed, indicating that the metadata was selectively modified to reflect the intended purpose.}
    %
\end{tcolorbox}
\vspace{-0.5em}


\PP{Replication by Different Maintainers (817 Packages)}
For replicated packages maintained by different accounts, the average name similarity with their original counterparts was {$0.56$} (median {$0.636$}; see \autoref{fig:name-cdf}). 
Among the {308} highly replicated packages, {149} ({48.37\%}) showed a name similarity score of 0.7 or higher, indicating that their names were often very similar. However, among the {509} partially replicated packages, {189} ({37.13\%}) reached a name similarity score of 0.7 or higher, suggesting a tendency to upload packages to PyPI with more than half of the code replicated but with significantly different names.
When a different maintainer republishes most of the code under a highly similar name, it may create user confusion, underscoring the need for careful monitoring to preserve ecosystem clarity and trust.


For metadata similarity, {12} packages had identical values across all five fields even though they were maintained by different accounts.
The number of replicated packages that directly reused elements of the original package's metadata is shown in \autoref{fig:meta}.
Overall, {198 (24.23\%)} highly and {205 (25.09\%)} partially replicated packages shared at least one identical field.

An interesting observation is that packages replicated by entirely different maintainers frequently retained the original summary. This may be due to convenience or an intentional choice to reduce the effort required while still benefiting from the credibility of the original package.
Note that metadata duplication was particularly prevalent among packages with high code similarity.
For example, \texttt{PyYAMLp-5.4.1} is nearly identical to \texttt{PyYAML-5.4.1}, sharing the same source code and all metadata fields, with only the package name and maintainer altered. Such cases are difficult to distinguish from the original and can mislead users.
Unlike the original projects, these replicated packages offer no guarantee of consistent maintenance and may retain unresolved vulnerabilities or introduce malicious changes. These findings emphasize the importance of scrutinizing replicated packages to maintain ecosystem trustworthiness.

\vspace{-0.3em}

\begin{tcolorbox}[colback=lime!13, 
    colframe=black, 
    boxrule=0pt, 
    left=5pt,
    right=5pt,
    top=2pt,
    bottom=2pt,
    enhanced jigsaw, 
    sharp corners]
    \textit{\textbf{Finding 5.} Despite being maintained by different accounts, {338} (out of {817; 39.58\%}) packages 
    exhibited name similarity scores above 0.7 with the original. We further observed that {403} replicated packages ({49.32\%}) 
    shared at least one identical metadata field, which increases the risk of user confusion and potential security threats.}
\end{tcolorbox}
\vspace{-0.5em}




\subsubsection{Code Similarity}

\begin{figure}[t]
	\begin{center}
			
			
        
		\begin{subfigure}[b]{0.44\textwidth}
			\centering
			\includegraphics[width=0.80\linewidth]{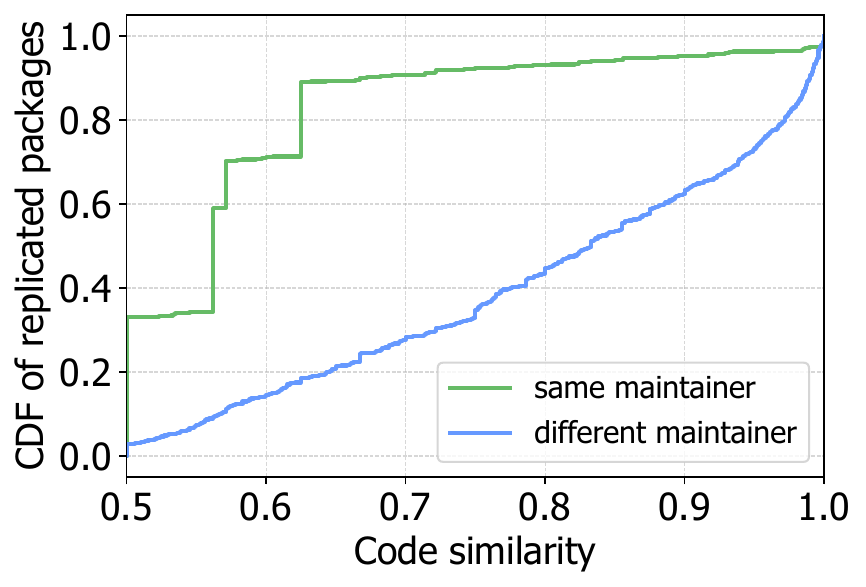}
	
			\vspace{-0.5em}
			\caption{Distribution of \textbf{code similarities} between original and replicated packages.}
			\label{fig:code-cdf}
		\end{subfigure}\hspace{1em}%
		\begin{subfigure}[b]{0.44\textwidth}
			\centering
			\includegraphics[width=0.80\linewidth]{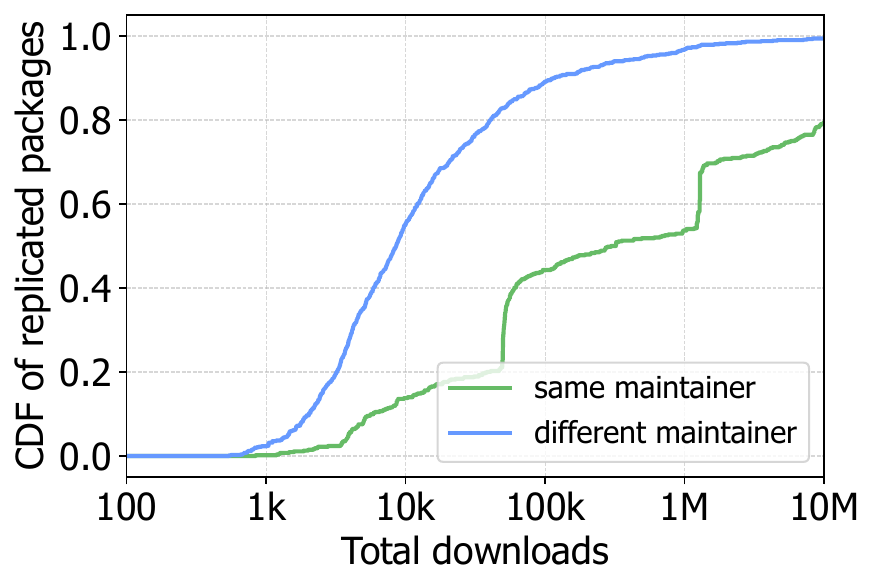}
	
			\vspace{-0.5em}
			\caption{Distribution of \textbf{total downloads} in replicated packages.}
			\label{fig:downloads-cdf}
		\end{subfigure}
		\vspace{-0.7em}
  
		\caption{CDF graphs of code similarity and total downloads. In general, same-maintainer replications exhibit higher code similarity and download counts.}\label{fig:cdfs}
	\end{center}
	
	\vspace{-1.0em}
\end{figure}

Next, we examine the code similarity between replicated packages and their original counterparts. 
\autoref{fig:code-cdf} shows the CDF of code similarity. Replicated packages with the same maintainer achieved an average code similarity score of {0.58} (median 0.56), whereas those with different maintainers {showed a higher} average of {0.80 (median 0.83)}.
\autoref{table:filenum} presents the types of file changes identified during replication. 
%
We observed that most files remained identical, while a smaller portion were modified, added, or deleted. On average, {85.04\%} of the original package's files were directly reused, {12.46\%} were replicated with modifications, and {2.50\%} were deleted. 
Although 
the proportion of modified files was modest,
the average line-level code modification ratio was approximately {39\%}, which implies that code changes were {non-trivial} once modifications occurred.

In addition, 
we observed that {32,310} files were newly added during replications. Although many appear to support functional extension, the possibility of security-relevant functionality calls for close scrutiny.
Moreover, replications by the same maintainer involved an average modification ratio of {0.53}, whereas those by different maintainers showed {lower} changes ({0.26}), {suggesting substantially less modification among different maintainers.}

\begin{table}[t]
\centering
\renewcommand{\tabcolsep}{3mm}	
	\caption{\label{table:filenum}Number of the files by type.}
    \vspace{-1em}
    \small	
\begin{tabular}{|c|c|c|c|c|c|}
\hline
\multicolumn{1}{|c|}{\rule{0in}{2.2ex}\textbf{Type}}
& \multicolumn{1}{c|}{Total}
& \multicolumn{1}{c|}{Identical}
& \multicolumn{1}{c|}{Modified}
& \multicolumn{1}{c|}{Added}
& \multicolumn{1}{c|}{Deleted}\\\hline\hline
\rule{0in}{2.2ex}\textbf{\#Files}
& {227,380}
& {165,887}
& {24,313}
& {32,310}
& {4,870}
\\\hline
\end{tabular}
\vspace{-1.0em}
\end{table}

\vspace{-0.3em}
\begin{tcolorbox}[colback=lime!13, 
    colframe=black, 
    boxrule=0pt, 
    left=5pt,
    right=5pt,
    top=2pt,
    bottom=2pt,
    enhanced jigsaw, 
    sharp corners]
    \textit{\textbf{Finding 6.} Replicated packages tend to reuse most of the original code: on average, {85.04\%} of the original files are carried over unchanged. When edits are made, changes are concentrated in a small subset of original files, with line-level modification ratios averaging {39\%}.}
    %
\end{tcolorbox}
\vspace{-0.5em}




\subsubsection{Popularity of Replicated Packages}
Finally, we investigated the popularity of replicated packages by analyzing download counts.
\autoref{fig:downloads-cdf} shows the CDF for download counts. 
Out of the {544} replicated packages maintained by the same maintainer, {303 (55.70\%)} had more than 100,000 downloads, indicating high popularity. Notably, packages with more than 10,000 downloads accounted for {469 (86.21\%)}.
Packages replicated by different maintainers generally exhibited lower download counts; however, several achieved substantial popularity. Out of {817} such packages, {89 (10.89\%)} recorded more than 100,000 downloads, {365 (44.67\%)} surpassed 10,000 downloads, and {26} packages {(3.18\%)} exceeded 1 million downloads (the most downloaded reached over 21 million).
\vspace{-0.3em}
\begin{tcolorbox}[colback=lime!13, 
    colframe=black, 
    boxrule=0pt, 
    left=5pt,
    right=5pt,
    top=2pt,
    bottom=2pt,
    enhanced jigsaw, 
    sharp corners]
    \textit{\textbf{Finding 7.} 
    Replicated packages have also gained notable popularity in the Python ecosystem, with many recording over 10,000 downloads. This suggests that users may inadvertently install these packages, sometimes confusing them with the original ones, which in turn can introduce unnecessary maintenance challenges or even potential security risks.}
    %
\end{tcolorbox}
\vspace{-0.5em}

\section{Vulnerability Analysis (RQ2)}\label{sec:rq2}
To understand the security implications of replication, we then examined how replicated packages overlap with known vulnerabilities. 
{For broad identification of vulnerable replicated packages, we used the vulnerability dataset as a reference and compared it against the candidate dataset pool.}

    \subsection{Status of Vulnerable Replications}
By comparing 1,072 vulnerable packages (as a reference) with the candidate pool, we initially identified {543} replicated packages.
However, some of the identified replications originated from patched versions rather than vulnerable ones. 
{To refine the initial results, we applied a two-step refinement process. First, through version-based inspection, we verified whether the duplicated package versions fell within the vulnerable version range defined by OSV. Next, through code-level analysis, we manually examined whether the security patches had been applied by comparing the patch code with the corresponding reused code.}
Finally, we identified \textbf{{256} packages} in which the vulnerable code had been directly replicated ({93} highly and {163} partially replicated packages).

\PP{Ethical Disclosure} We prioritized verification starting from the most-downloaded packages and reported vulnerable packages in which the vulnerabilities could be triggered. Unlike malicious packages (see \autoref{subsec:wild}), however, our vulnerability reports to PyPI (or the corresponding team) often received no response, or in some cases, only an acknowledgment without a subsequent patch. To date, we have reported 10 vulnerabilities and received confirmation for only two cases. We plan to continue reporting vulnerabilities that require immediate attention on an ongoing basis.

\vspace{-0.3em}
\begin{tcolorbox}[colback=lime!13, 
    colframe=black, 
    boxrule=0pt, 
    left=5pt,
    right=5pt,
    top=2pt,
    bottom=2pt,
    enhanced jigsaw, 
    sharp corners]
    \textit{\textbf{Finding 8.} From 1,072 vulnerable packages, we identified {256} replications that directly preserved the vulnerability, underscoring the security risks posed by vulnerable replications in the ecosystem.}
\end{tcolorbox}
\vspace{-0.7em}

\subsection{Characteristics of Replicated Vulnerable Packages}




\begin{figure}[t]
	\begin{center}
         \begin{subfigure}[b]{0.44\textwidth}
			\centering
                \includegraphics[width=0.85\linewidth]{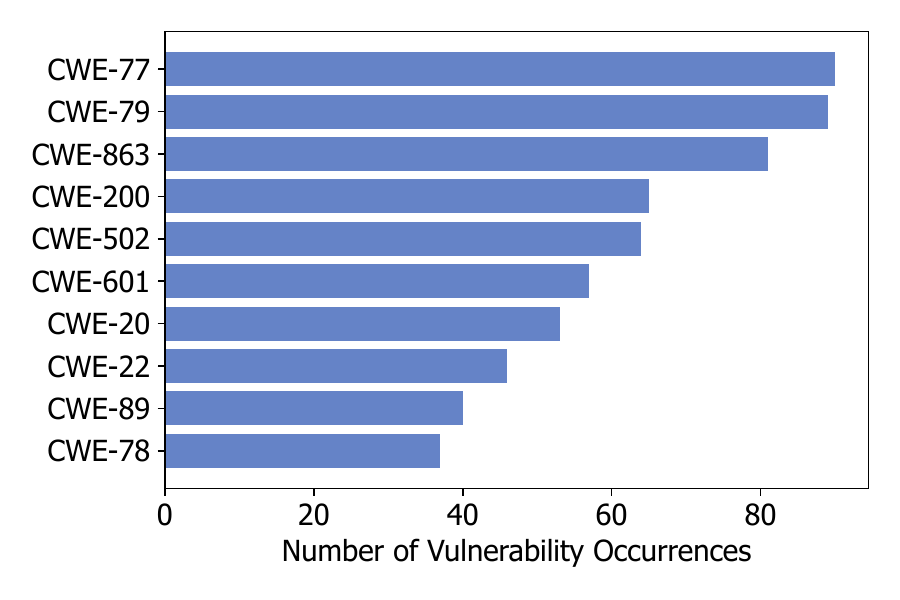}
			
			\vspace{-0.3em}
			\caption{Top 10 \textbf{CWE}s by vulnerability count.}
			\label{fig:cwe}
            \vspace{0.5em}
		\end{subfigure}\hspace{1em}%
		\begin{subfigure}[b]{0.44\textwidth}
			\centering
                \includegraphics[width=0.85\linewidth]{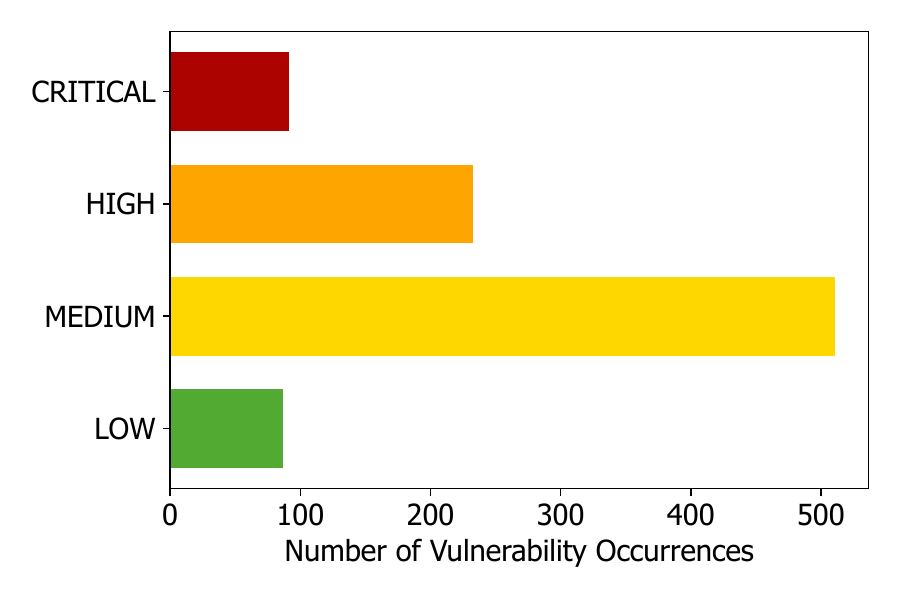}
			
			\vspace{-0.3em}
			\caption{\textbf{CVSS} distribution of vulnerabilities.}
			\label{fig:severity}
            \vspace{0.5em}
		\end{subfigure}
        \vspace{-1.0em}
		\caption{Graphs showing the vulnerability types and severities of the identified replicated vulnerable packages.} 
        \label{fig:cve}
	\end{center}
	
	\vspace{-1.3em}
\end{figure}

\subsubsection{Vulnerability Types and Distribution}
Among the {256} identified replications, we observed a total of {1,025} vulnerability occurrences and {383} unique vulnerabilities  
(comprising {370} CVE-listed vulnerabilities and {13} additional issues without CVE identifiers).
On average, each package contained four vulnerabilities, indicating that replicated packages often inherited multiple flaws.

To analyze vulnerability types and severity, we use Common Weakness Enumeration (CWE) and Common Vulnerability Scoring System (CVSS). Several non-CVE vulnerabilities also provide this information through PYSEC and GHSA, thus, this decision enables a consistent analysis.

The distribution of vulnerabilities by type and severity is summarized in \autoref{fig:cve}. 
Regarding vulnerability types, input validation-related vulnerabilities  (\eg, CWE-20: Improper Input Validation and CWE-77: Command Injection) were the most common, as they frequently arise from developers mishandling untrusted user data. This was followed by access control failures (\eg, CWE-863: Improper Authorization), which often result from insufficient restrictions on sensitive data. Redirection and deserialization issues (\eg, CWE-601: Open Redirect and CWE-502: Deserialization of Untrusted Data) were also highly represented, particularly in Python due to the common use of built-in serialization libraries and the need for careful path handling.

In terms of severity, medium-severity vulnerabilities ({{511}} cases) were the most prevalent, followed by high ({233}), critical ({91}), and low ({87}) severities. This finding suggests that replicated packages frequently contain vulnerabilities of substantial risk, underscoring the need for timely mitigation.
In particular, high and critical vulnerabilities increase the possibility of exploitation and, if reused as dependencies, may propagate downstream, amplifying their impact.

\vspace{-0.3em}
\begin{tcolorbox}[colback=lime!13, 
    colframe=black, 
    boxrule=0pt, 
    left=5pt,
    right=5pt,
    top=2pt,
    bottom=2pt,
    enhanced jigsaw, 
    sharp corners]
    \textit{\textbf{Finding 9.} The {256} identified replication packages contained {383} unique vulnerabilities ({370} CVEs), with input validation flaws being most common, followed by access control and data handling issues. In addition, we found that high- and critical-severity vulnerabilities account for {35.14\%}, which indicates that replicated vulnerable packages require more careful attention and mitigation.} 
\end{tcolorbox}
\vspace{-0.5em}



\subsubsection{Maintenance Aspects of Replicated Vulnerable Packages}
\begin{figure}[t]
	\begin{center}
			
			
        
		\begin{subfigure}[b]{0.44\textwidth}
			\centering
			\includegraphics[width=0.85\linewidth]{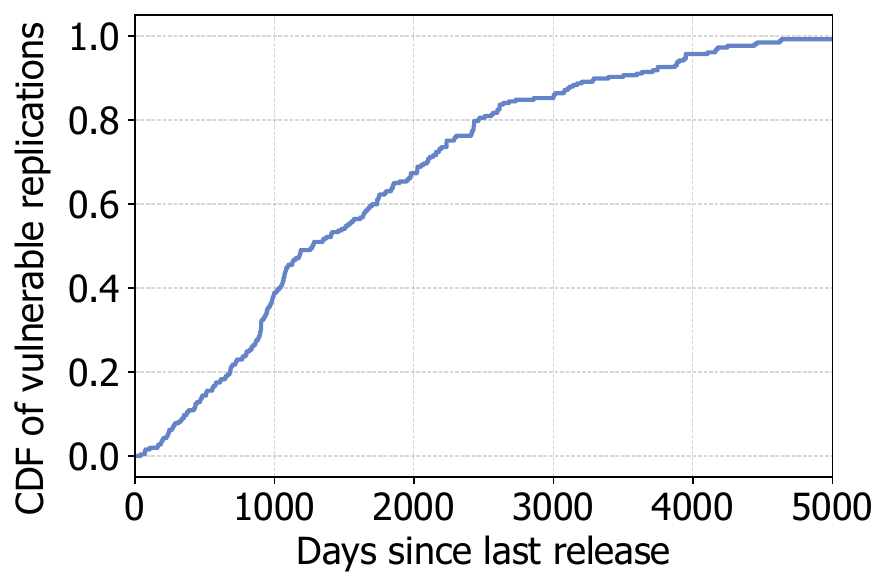}
	
			\vspace{-0.3em}
			\caption{Distribution of \textbf{days since last release} of vulnerable replication.}
			\label{fig:lastrelease}
		\end{subfigure}\hspace{1em}%
		\begin{subfigure}[b]{0.44\textwidth}
			\centering
			\includegraphics[width=0.85\linewidth]{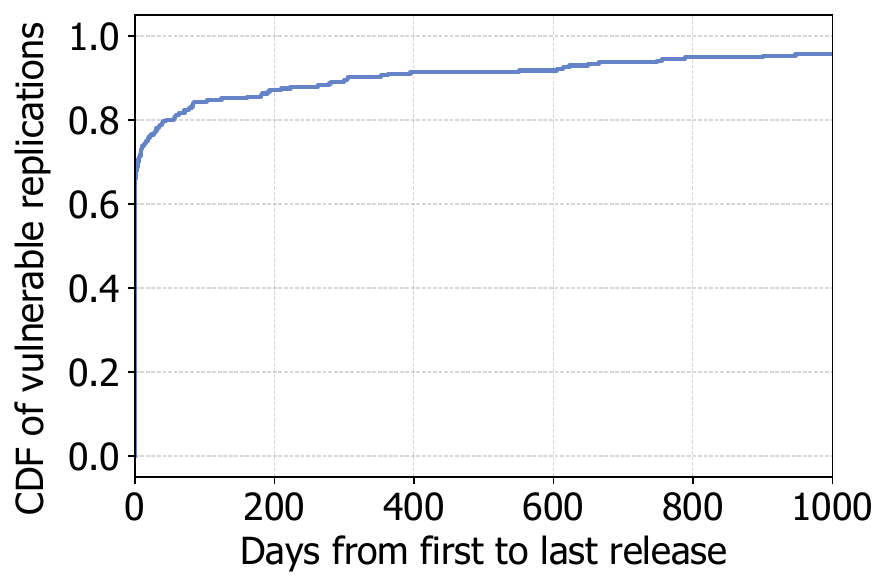}
	
			\vspace{-0.3em}
			\caption{Distribution of \textbf{days from first to last release} of vulnerable replication.}
			\label{fig:lifetime}
		\end{subfigure}
		\vspace{-0.7em}
  
		\caption{CDF graphs of days since last release and days from first to last release. 
        }\label{fig:maintenance}
	\end{center}
	
	\vspace{-1.3em}
\end{figure}

To assess the maintenance level of the replicated vulnerable packages, we analyzed their release activity. \autoref{fig:maintenance} shows the maintenance status of vulnerable replications.
Among the {256} vulnerable packages, {182} ({71.09\%}) exhibited a release interval of less than one week between their initial release and the most recent update.
In addition, {231} ({90.23\%}) out of the {256} packages had not been updated for over a year as of September 2025.
The fact that the last update of the replicated vulnerable packages dates back approximately {1,630} days on average (median {1,274} days) suggests that they are either poorly maintained or left unattended after release.
Notably, {92} of these poorly maintained packages had more than 10,000 downloads, indicating that users continued to install them despite their outdated and insecure state.




\vspace{-0.3em}
\begin{tcolorbox}[colback=lime!13, 
    colframe=black, 
    boxrule=0pt, 
    left=5pt,
    right=5pt,
    top=2pt,
    bottom=2pt,
    enhanced jigsaw, 
    sharp corners]
    \textbf{\textit{Finding 10.}}
    \textit{Of the {256} vulnerable package replications, 
    {231} ({90.23\%})
    were unmaintained for over a year,
    with an average of {1,630} days since their last update. 
    This suggests that most of these packages remain effectively unmaintained.}
    %
\end{tcolorbox}
\vspace{-0.5em}

\subsubsection{Detection Gaps in Existing Tools}
{Next, we evaluated the detectability of vulnerable replications identified by our replication-based approach. Specifically, we used \mbox{\texttt{DependencyTrack}}~\mbox{\cite{dptrack}} and \texttt{Safety}~\mbox{\cite{safety}}, two widely adopted tools detecting vulnerable packages via dependency metadata.}

{Out of the 256 vulnerable replications identified, only four and one cases were detected by \texttt{Safety} and \texttt{DependencyTrack}, respectively. This discrepancy arises because dependency-based scanners rely on known package metadata, such as names and versions. Consequently, repackaged vulnerable code under different identifiers can evade detection.}
For example, \autoref{lst:vulcode} shows a vulnerability propagated through replication (for ethical reasons, we show only the original CVE code). The flaw, originally reported in \texttt{Gradio}, enables arbitrary file copying and may lead to denial-of-service attacks. We found a replicated package containing the same vulnerable code without modification. Because it was published under a different name and versioning scheme, conventional dependency-based techniques failed to detect it.
In contrast, our code-based replication analysis successfully uncovered such vulnerable replicas that existing approaches overlook.

\vspace{-0.4em}
\begin{tcolorbox}[colback=lime!13, 
    colframe=black, 
    boxrule=0pt, 
    left=5pt,
    right=5pt,
    top=2pt,
    bottom=2pt,
    enhanced jigsaw, 
    sharp corners]
    \textit{\textbf{Finding 11.} Dependency-based vulnerability scanners fail to detect replicated vulnerable packages because they rely on original package information. Replicated packages distribute the same vulnerable code under different names, creating invisible security risks.}
    %
\end{tcolorbox}
\vspace{-1.5em}

\begin{lstlisting}[float, caption={\label{lst:vulcode}Example of propagated vulnerable code (CVE-2025-48889).}, 
language=C, frame=single, style=base, frame=btlr, 
basicstyle=\fontsize{7}{8}\ttfamily\color{black}, 
deletekeywords={for}, mathescape, breakatwhitespace=true, breaklines=true,
numbers=left, numbersep=8pt, xleftmargin=1em]
$\text{\;\;\textcolor{black}{def \_copy\_to\_dir(self, dir: str) -> FileData:}}$
$\text{\;\;\quad\textcolor{black}{pathlib.Path(dir).mkdir(exist\_ok=True)}}$
$\text{\;\;\quad\textcolor{black}{new\_obj = dict(self)}}$
$\text{\;\;\quad\textcolor{black}{if not self.path:}}$
$\text{\;\;\quad\quad\textcolor{black}{raise ValueError("Source file path is not set")}}$
$\text{\;\;\quad\textcolor{black}{new\_name =} \colorbox{fsepink!30}{shutil.copy}(self.path, dir) \textcolor{gray}{\# vulnerable sink}}$
$\text{\;\;\quad\textcolor{black}{new\_obj["path"] = new\_name}}$
$\text{\;\;\quad\textcolor{black}{return self.\_\_class\_\_(**new\_obj)}}$
\end{lstlisting}
\setlength{\textfloatsep}{0pt}
\setlength{\intextsep}{0pt}
\setlength{\floatsep}{1pt}

\section{Malware Analysis (RQ3)}\label{sec:rq3}
Finally,
we examine how package replication contributes to malicious code distribution.

\subsection{Analysis of Known Malware Exploiting Package Replication}

\subsubsection{Status of Replicated Malicious Packages}
To examine how replication is exploited in real-world attacks, we compared the popularity dataset against the malware dataset (\ie, 2,656 known malicious packages).
We confirmed that {186 (4.79\%)} known malicious packages were replicated from popular packages: {125} highly and {61} partially replicated packages.
Replicating code from popular packages increases the possibility of evading code-based malware detection and exacerbates developer confusion, potentially leading to inadvertent malware installation and broader propagation.
Notably, most cases replicated over 90\% of the original code, indicating minimal modification by attackers and highlighting package replication as a notable vector for malware distribution.
%

\vspace{-0.5em}
\begin{tcolorbox}[colback=lime!13, 
    colframe=black, 
    boxrule=0pt, 
    left=5pt,
    right=5pt,
    top=2pt,
    bottom=2pt,
    enhanced jigsaw, 
    sharp corners]
    \textit{\textbf{Finding 12.} {186 (4.79\%)} of known malicious packages were replications of popular packages, a majority (67\%) being highly replicated ($\geq$90\% code reuse), demonstrating that attackers exploit package replication as an {observable} distribution mechanism by making minimal modifications.}
    %
\end{tcolorbox}
\vspace{-0.9em}


\setlength{\textfloatsep}{0pt}
\setlength{\intextsep}{0pt}
\begin{figure}[t]
	\begin{center}
         \begin{subfigure}[b]{0.35\textwidth}
			\centering
                \includegraphics[width=0.90\linewidth]{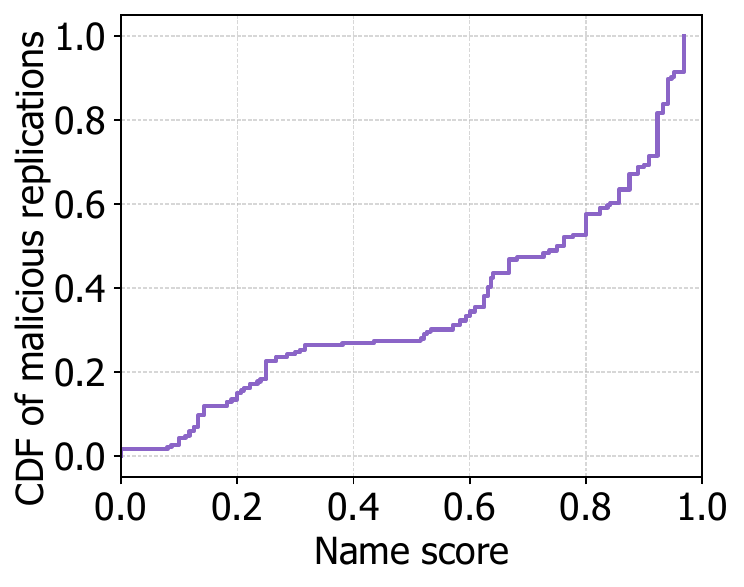}
			
			\vspace{-0.3em}
			\caption{Name similarity between replicated malicious packages and their originals.}
			\label{fig:mal_name}
		\end{subfigure}\hspace{2em}%
		\begin{subfigure}[b]{0.5\textwidth}
			\centering
                \includegraphics[width=0.90\linewidth]{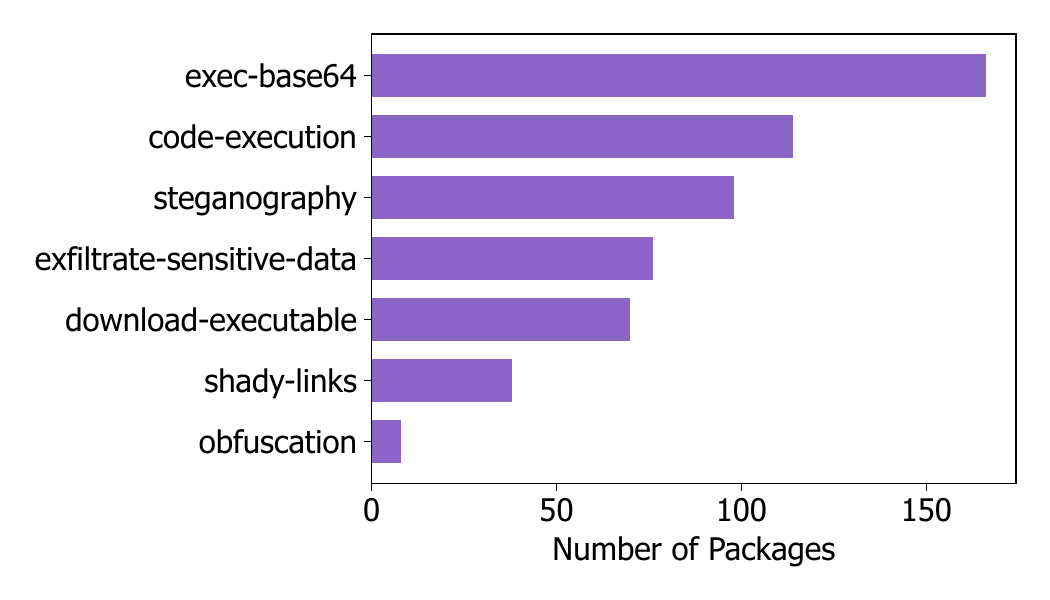}
			
			\vspace{-0.3em}
			\caption{Statistics of suspicious APIs added in replicated malicious packages.}
			\label{fig:mal_sus}
		\end{subfigure}
        \vspace{-0.5em}
		\caption{Graphs of name similarity in replicated malicious packages and statistics of suspicious APIs.}\label{fig:mal_statistic}
	\end{center}
	
	\vspace{0.5em}
\end{figure}

\subsubsection{Name and Metadata Similarity}
\autoref{fig:mal_name} shows the CDF of name similarities for  replicated malicious packages.
Notably, 
{88 (47.31\%)} out of {186} had a name similarity score of 0.8 or higher, often reflecting typosquatting patterns (\eg, \texttt{jeilyfish} mimicking \texttt{jellyfish}).
%
However, some malicious packages avoided typosquatting by replicating most of the code under entirely different names. For instance, \texttt{youtube-new} replicated \texttt{requests}, and \texttt{u283udsfru} replicated \texttt{certifi}. In such cases, name-based detection is ineffective, highlighting the necessity of code-based identification.


Next, we measured the metadata similarity.
Because the description field was unavailable for known malicious packages, similarity was computed using the remaining four fields. 

We observed that metadata replication was widespread.
Among {186} cases, {159} ({85.48\%}) shared at least one field with their originals. Of the {125} highly replicated packages, {108} ({86.4\%}) duplicated at least one field and {66} ({52.8\%}) duplicated three or more. Partially replicated packages showed a similar but weaker pattern, with {50} ({81.96\%}) sharing at least one field and {32} ({52.45\%}) sharing three or more.
The most frequently copied field was \texttt{summary} ({119}), followed by \texttt{author\_email} ({109}), \texttt{home\_page} ({99}), and \texttt{author} ({73}). These results indicate that many malicious packages replicate not only code but also identifying metadata to enhance credibility or evade detection.

\vspace{-0.3em}
\begin{tcolorbox}[colback=lime!13, 
    colframe=black, 
    boxrule=0pt, 
    left=5pt,
    right=5pt,
    top=2pt,
    bottom=2pt,
    enhanced jigsaw, 
    sharp corners]
    \textit{\textbf{Finding 13.} Among the replicated malicious packages, {88 (47.31\%)} showed high name similarity ($\geq$0.8) with their originals and {158 (84.94\%)} shared at least one metadata field, indicating that attackers frequently copy both names or metadata to mislead users and complicate detection.}
\end{tcolorbox}
\vspace{-0.5em}

\subsubsection{Suspicious API Injection}\label{subsubsec:sus}
Next, we investigated the additional functionalities in replicated malicious packages by focusing on the modified and newly added files (see \autoref{subsubsec:code_sim}).

Our analysis revealed that {180 (96.77\%)} out of {186} replicated malicious packages incorporated new APIs absent from the original repositories.
Specifically, we identified {796} unique APIs newly introduced across the {180} replicated malicious packages. These additions frequently appeared in combination, with multiple suspicious APIs embedded within a single line of code.




To conduct a clear analysis of these APIs, we leveraged \texttt{GuardDog}'s \texttt{Semgrep} ruleset~\cite{guarddog}. Among the ruleset, seven categories of suspicious behaviors were detected.

\begin{enumerate}
    \setlength\itemsep{0.2em}
    \item \textbf{\texttt{Shady links.}} Embedding or redirecting to suspicious or malicious external websites. 
    \item \textbf{\texttt{Obfuscation.}} Hiding code logic through encoding, packing, or complex transformations. 
    \item \textbf{\texttt{Download executable.}} Fetching and saving external binary files that may contain malware. 
    \item \textbf{\texttt{Exfiltrate sensitive data.}} Collecting and transmitting private information.
    \item \textbf{\texttt{Steganography.}} Concealing malicious code or data within seemingly benign files or media. 
    \item \textbf{\texttt{Code execution.}} Executing arbitrary or attacker-controlled code on the victim system. 
    \item \textbf{\texttt{Exec-base64.}} Executing Base64-encoded payloads to bypass simple inspection. 
\end{enumerate}

Following their definitions, we examined additional suspicious APIs introduced in replicated malicious packages.
As shown in \autoref{fig:mal_sus}, the most prevalent type was \texttt{exec-base64}, appearing {582} times across {166} packages, as it enables payload decoding and execution in a single step.
\texttt{Code execution} APIs were added {217} times in {114} packages, allowing direct command or script execution.
\texttt{Steganography} appeared {173} times in {98} packages, facilitating concealment of malicious content.
We also identified APIs related to \texttt{exfiltrate sensitive data} ({76} packages), \texttt{download executable} ({70}), \texttt{shady links} ({38}), and \texttt{obfuscation} ({8}).

Beyond single API insertion, some replicated malicious packages leveraged combinations of these suspicious behaviors. For example, {37} replicated packages were found to include APIs related to \texttt{steganography}, \texttt{code} \texttt{execution}, and \texttt{exec-base64} simultaneously.

\setlength{\textfloatsep}{4pt}
\begin{tcolorbox}[colback=lime!13, 
    colframe=black, 
    boxrule=0pt, 
    left=5pt,
    right=5pt,
    top=2pt,
    bottom=2pt,
    enhanced jigsaw, 
    sharp corners]
    \textit{\textbf{Finding 14.} {180} replicated malicious packages {(96.77\%)} injected {796} unique suspicious APIs, with exec-base64 and code execution being most common, and these suspicious functionalities often appeared in combination, suggesting sophisticated malicious intent.}
\end{tcolorbox}
\vspace{-0.5em}


\vspace{0.1em}
\subsubsection{Case Study}
We present a case in which package replication was used to introduce malicious code.
We identified \texttt{python-aliyun-sdk-core-2.13.36} as a highly replicated version of \texttt{aliyun-python-sdk-core-2.13.36}, with a code similarity score of {0.993}. The package names were highly similar, and all four metadata fields were identical.
Of the original package’s {146} files, {144} were reused without modification, while two files, \texttt{aliyunsdkcore/client.py} and \texttt{setup.py}, were altered. In particular, malicious logic was injected into the \texttt{\_resolve\_endpoint} function in \texttt{client.py}, as shown in \autoref{lst:malcode1}.
This function determines service endpoints based on region and product information. In the replicated version, the attacker introduced suspicious APIs related to \texttt{exfiltrate sensitive data} and \texttt{shady links}, attempting to leak access keys such as \texttt{ak} and \texttt{secret} to an external domain. This example illustrates a credential exfiltration pattern embedded within an otherwise near-identical replica of a legitimate SDK.

\begin{lstlisting}[float, caption={\label{lst:malcode1}Malicious code injected in the \texttt{\_resolve\_endpoint} function.}, 
language=C, frame=single, style=base, frame=btlr, 
basicstyle=\fontsize{7}{8}\ttfamily\color{black}, 
deletekeywords={for}, mathescape, breakatwhitespace=true, breaklines=true,
numbers=left, numbersep=8pt, xleftmargin=1em]
$\text{\quad\textcolor{black}{  resolve\_request.endpoint\_regional = request.endpoint\_regional}}$
$\text{\colorbox{fsegreen!15}{+ try:}}$
$\text{\colorbox{fsegreen!15}{+\;\;\quad sessions = Session()}}$
$\text{\colorbox{fsegreen!15}{+\;\;\quad data = \{"ak": self.\_ak, "secret": self.\_secret\}}}$
$\text{\colorbox{fsegreen!15}{+\;\;\quad sessions.close()}}$
$\text{\colorbox{fsegreen!15}{+ except:}}$
$\text{\colorbox{fsegreen!15}{+\;\;\quad pass}}$
$\text{\quad\textcolor{black}{ return self.\_endpoint\_resolver.resolve(resolve\_request)}}$
\end{lstlisting}



\vspace{-0.3em}
\subsection{Replicated Malicious Package in the Wild}\label{subsec:wild}


We applied our replication detection and suspicious API injection analysis to the recent dataset collected between May and August 2025, comparing newly uploaded packages against the popularity dataset. 
{We used the recent dataset rather than the candidate dataset to enable early detection and timely response to potentially malicious code introduced through new package releases. Malware is often removed or modified shortly after upload, so recent data more accurately reflects attacker activity. Accordingly, prior Python malware-detection research has also focused on recently uploaded packages (\eg, \mbox{\cite{maloss, malwukong}}), rather than on randomly sampled data.}



{Malicious package detection was conducted following the approach described in \mbox{\autoref{subsubsec:sus}}. Specifically, we tracked recent packages that replicated popular packages and manually examined whether they contained APIs associated with malicious behavior.}

From a recent dataset of 25,407 packages, we identified {227} ({0.89\%}) that replicated popular packages. Among them, {67} contained previously identified suspicious APIs in newly added or modified files.
We manually inspected these {67} cases, focusing on code regions that differed from the originals, and confirmed malicious behaviors such as data exfiltration and unauthorized downloads. As a result, we uncovered \textbf{seven previously unknown malicious packages}, all of which were removed from PyPI after our disclosure.
One representative case, \texttt{zoz-requests}, replicated \texttt{requests} with minimal changes but injected code into the \texttt{request} function to redirect traffic through a hardcoded proxy and disable SSL verification, enabling man-in-the-middle attacks (see \autoref{lst:malcode2}). The package remained on PyPI for 123 days before removal, demonstrating how malicious replicas can evade detection by blending into trusted codebases.

\vspace{-0.3em}
\begin{tcolorbox}[colback=lime!13, 
    colframe=black, 
    boxrule=0pt, 
    left=5pt,
    right=5pt,
    top=2pt,
    bottom=2pt,
    enhanced jigsaw, 
    sharp corners]
    \textit{\textbf{Finding 15.} Our replication- and suspicious API–based analysis uncovered seven unknown malicious packages, confirming that replication remains an attack vector in recent uploads.} 
\end{tcolorbox}

\begin{lstlisting}[float, caption={\label{lst:malcode2}Malicious code injected in the \texttt{request} function.}, 
language=Python, frame=single, style=base, frame=btlr, 
basicstyle=\fontsize{7}{8}\ttfamily\color{black}, 
mathescape, breakatwhitespace=true, breaklines=true,
numbers=left, numbersep=8pt, xleftmargin=1em]
$\text{\quad\textcolor{black}{def request(method, url, **kwargs):}}$
$\text{\quad\;\;\;\textcolor{black}{with sessions.Session() as session:}}$
$\text{\colorbox{fsegreen!15}{+\;\;\;\quad kwargs.setdefault("proxies", {"http": "http://192.168.0.128:8080", "https":"http://192.168.0.128:8080"})}}$
$\text{\colorbox{fsegreen!15}{+\;\;\;\quad kwargs.setdefault("verify", False)  \# Bypass SSL verification}}$
$\text{\quad\;\;\;\;\;\textcolor{black}{return session.request(method=method, url=url, **kwargs)}}$
\end{lstlisting}
\setlength{\floatsep}{4pt}


\setlength{\textfloatsep}{7pt}
\section{Discussion}

\subsection{Similarity Threshold Sensitivity}\label{subsec:simsen}
{\mbox{\autoref{table:disstribution_res}} summarizes the distribution of similarity scores across the RQs. For RQ2, we report the number of replicated packages that copied vulnerable versions. For RQ3, we exclude explicit false positives, such as cases involving only test file replication. As expected, lowering the threshold identifies more replicated packages but also introduces more false positives. Notably, the number of packages in the 0.9-1.0 range exceeds those in the 0.8-0.9 range, supporting 0.9 as a meaningful boundary for highly replicated packages. Similarly, the 0.5 threshold effectively separates partially replicated packages from lower-similarity cases in RQ1 and RQ3.}
{Although both thresholds were empirically chosen, they can be adjusted based on analytical objectives (\eg, a lower threshold may be preferred when recall is prioritized). Overall, the observed distributions suggest that our chosen thresholds provide a reasonable and well-balanced criterion for replication analysis.}
\vspace{-0.5em}

\begin{table}[t]
\centering
\renewcommand{\tabcolsep}{1mm}	
	\caption{\label{table:disstribution_res}{Distribution of replicated package detection results by the similarity score.}}
    \vspace{-1.3em}
    \small
	\begin{center}	
\begin{tabular}{|c||c|c|c|c|c|c|c|c|c|c||c|}
\hline
\rule{0in}{2.2ex}\textbf{Similarity score}
& 0.0-0.1
& 0.1-0.2
& 0.2-0.3
& 0.3-0.4
& 0.4-0.5
& 0.5-0.6
& 0.6-0.7
& 0.7-0.8
& 0.8-0.9
& 0.9-1.0
& Total\\\hline\hline
\rule{0in}{2.2ex}\textbf{RQ1 (D1 vs. D5)}
& 3,114 & 572 & 536 & 315 & 177 & 575 & 229 & 141 & 166 & 375 & 6,200\\\hline
\rule{0in}{2.2ex}\textbf{RQ2 (D2 vs. D5)}
& 1,259 & 152 & 79 & 54 & 40 & 39 & 48 & 73 & 68 & 129 & 1,977\\\hline
\rule{0in}{2.2ex}\textbf{RQ3 (D1 vs. D4)}
& 13 & 5 & 7 & 8 & 10 & 22 & 20 & 11 & 13 & 126 & 235\\\hline
\end{tabular}
\end{center}
\end{table}
\vspace{0.3em}

\subsection{Application}


\PP{Application to Code Clone Detection}
Our methodology for identifying packages that replicate most of the original code can be applied to code clone detection research in package manager ecosystems. In particular, our framework integrates embedding, clustering, code similarity, and name/metadata similarity rather than relying on any single signal, offering useful intuition for analyzing code reuse across various ecosystems, including Python.

\PP{Application to Vulnerability Detection}
The discovery of vulnerable replications underscores the need for code-level vulnerability detection in PyPI. Prior 1-day vulnerability studies targeting package manager ecosystems relied on package names, versions, and dependency information. By contrast, vulnerability analysis for C/C++, where code-level reuse is more prevalent, has depended on direct code analysis (\eg, \cite{woo2022movery, feng2024fire}). Our findings suggest that code-level analysis is equally necessary for PyPI, and provide a foundation for future work on vulnerable packages.


\PP{Application to Malware Detection}
Our findings suggest a new research direction for detecting malicious packages that prior approaches have not fully addressed. Existing PyPI malware detection studies have focused on code shared across packages—identifying those with functionalities similar to known malware (\eg, \cite{mphunter, malguard}). In contrast, our findings highlight the importance of detecting malicious packages by focusing on subtle \textit{differences} within highly similar packages. Statistical analysis of suspicious APIs can provide a basis for such detection and foster future research.




\subsection{Replication in Open-Source Ecosystems}
{Open-source ecosystems encourage reuse and redistribution, and package replication can arise for legitimate reasons such as customization or compatibility maintenance. Our study does not treat replication as inherently problematic. Rather, we focus on its security implications: replication can be exploited for malicious purposes, and it can also unintentionally preserve vulnerabilities when patches are not consistently propagated. By examining replication from a security perspective, we aim to understand how this common engineering practice may introduce risks at ecosystem scale.}

\subsection{Limitations}
First, our detection pipeline focused exclusively on Python source files, excluding other components (\eg, C extensions). Although manual inspection confirmed the presence of some vulnerabilities in such files, our automated replication detection was not applied beyond Python. 
Second, in some cases the accuracy of clone detection decreases due to the small size of packages and environment leakage (\eg, site-packages included in distributions).
{Third, manual analysis remains essential for confirming both malicious and vulnerable packages. Although our framework aims to identify replication and inherited security risks, false negatives may arise due to limitations in embedding, clustering, or similarity thresholds, potentially leading to missed replications or undetected vulnerable variants.}
Lastly, the scope of this study is limited to vulnerabilities and malicious activities arising from package cloning. Other issues, such as vulnerabilities introduced through insecure coding practices, are out of the scope of this study.

\subsection{Threats to Validity}
First, our dataset represents approximately one-third of PyPI packages, which may not fully capture the characteristics of the entire Python ecosystem. 
{Second, the distinction between vulnerable and patched versions can be subtle, as patches often introduce only minor code changes. Although we perform manual inspection to analyze the relevant code regions and assess whether vulnerability-inducing logic persists, the mere presence of vulnerable patterns does not guarantee exploitability in practice. Consequently, some flagged cases may not be practically exploitable under realistic attack conditions, and conversely, certain exploitable cases may remain undetected.}
{Third, as our analysis focuses on replicated packages that reuse a substantial portion of an existing codebase, malicious variants that incorporate only small code fragments fall outside our scope. Nevertheless, our approach is complementary to existing techniques that detect malicious packages based on fine-grained code reuse or small injected snippets.}
Finally, parts of the study relied on manual inspection by experts. Although multiple reviewers were involved in the investigation process to enhance reliability, this manual component inevitably introduces subjectivity and potential inconsistencies. The interpretation of vulnerability persistence and suspicious behavior may vary between reviewers, potentially affecting the reproducibility and generality of our findings.
\section{Related Work}

\PP{Code Clone Detection}
Prior research has mainly focused on detecting code clones, either from a general software engineering perspective or from a security perspective.
Early studies developed scalable techniques to detect similar code fragments across large codebases (\eg, \cite{jiang2007deckard, sajnani2016sourcerercc, nakagawa2021nil}),
while others aimed to identify vulnerable clones using pattern matching, machine learning, or function-level abstraction (\eg, \cite{jang2012redebug, li2016vulpecker, kim2017vuddy, woo2022movery, woo2023v1scan}).
More recently, researchers have shown that code cloning is pervasive across ecosystems, with large-scale studies reporting high duplication rates in GitHub and persistent vulnerabilities in shrinkwrapped clones in npm (\eg, \cite{lopes2017dejavu, wyss2022fork}).
These studies attempted to identify small reused code fragments or to analyze packages from a vulnerability perspective. However, they did not closely investigate the phenomenon of large-scale code replication, nor did they attempt to apply it to vulnerable or malicious packages.
Building on these insights, our work provides the first large-scale prevalence and security analysis of such replication in the Python ecosystem, revealing how it creates security risks through vulnerability preservation and malicious code injection.

\PP{Security Issue Detection in Python}
Previous research has explored various approaches to detecting malicious and vulnerable packages in PyPI. Early rule-based techniques combined static, dynamic, and metadata analysis (\eg, \cite{cho2025cryptbara, maloss, malwukong}), demonstrating practical detection capabilities but remaining limited against novel or obfuscated attacks. To overcome these limitations, machine learning has been increasingly adopted, using code embeddings, behavioral modeling, and graph-based analysis to improve detection accuracy (\eg, \cite{mphunter, ea4mp, malguard, cerebro}). Although they are effective, these methods require substantial training data and computational resources, and often focus on specific files (\eg, \texttt{setup.py}) or predefined behavioral patterns. \cite{vu2021lastpymile} compared package registry code against upstream repositories but they do not capture emerging malicious uploads.
{Typosquatting attacks have been widely studied in supply chain security through name-based similarity analysis (\eg, \mbox{\cite{vu2020typosquatting, kaplan2021survey, taylor2020spellbound, neupane2023beyond}}). These approaches focus on lexical characteristics of package identifiers and do not examine code-level replication across packages, which is our focus.}
ML-based approaches have also been proposed for detecting vulnerabilities (\eg, \cite{zhao2024python, wartschinski2022vudenc, mechri2025secureqwen}), but remaining limited in capturing complex contextual patterns. Moreover, analysis of commit histories shows that vulnerability fixes are often delayed \cite{antal2020exploring}.
In contrast to these approaches, our work shifts the focus from detecting suspicious patterns within individual packages to analyzing replication across the ecosystem, revealing how vulnerabilities and malicious code propagate through cloned distributions.

\section{Conclusion}

As PyPI has become increasingly central in the software ecosystem, ensuring the security of published packages has emerged as a critical concern. To address this, we conducted a large-scale empirical study on package replication, a key factor affecting package security. We identified {1,361} replicated packages and {256} replicated vulnerable packages, and further uncovered seven previously unknown malicious packages that had emerged through package replication.
Our findings highlight the characteristics and prevalence of package replication and, from a security perspective, show how it propagates vulnerabilities, enables suspicious modifications, and undermines ecosystem security. Our replication-based approach complements existing dependency- and metadata-driven tools, and in future work, we plan to develop security-oriented clone detection techniques and extend our pipeline beyond Python to other ecosystems.

\section*{Data Availability}
Our code and results are available at \url{https://github.com/sunha21/pypi-replication-analysis}.



\begin{acks}
This work was supported by the Institute of Information \& Communications Technology Planning \& Evaluation (IITP) grant funded by the Korea government (MSIT) (No.RS-2024-00440780, Development of Automated SBOM and VEX Verification Technologies for Securing Software Supply Chains), the National Research Foundation of Korea (NRF) grant funded by the Korea government (MSIT) (RS-2025-00517788, Research on Intelligent SBOM Generation and Automated Vulnerability Analysis through Multi-level Code Analysis),and the Culture, Sports and Tourism R\&D Program through the Korea Creative Content Agency grant funded by the Ministry of Culture, Sports and Tourism (International Collaborative Research and Global Talent Development for the Development of Copyright Management and Protection Technologies for Generative AI, RS-2024-00345025).
\end{acks}

\bibliographystyle{ACM-Reference-Format}
\bibliography{references}

\appendix



\end{document}